\documentclass[useAMS,usenatbib]{mn2e}
\usepackage{graphicx}

\title[Polarized Diffuse Emission at 2.3~GHz in a High Galactic Latitude Area]
      {Polarized Diffuse Emission at 2.3~GHz in a High Galactic Latitude Area}
\author[E. Carretti, et al.]
{E.~Carretti$^{1}$\thanks{E-mail:
carretti@bo.iasf.cnr.it}, 
D.~McConnell$^{2}$, 
N.~M.~McClure-Griffiths$^{2}$,
G.~Bernardi$^{1}$, \and
S.~Cortiglioni$^{1}$ and S.~Poppi$^{3}$\\
$^{1}$INAF--IASF Bologna, Via Gobetti 101, Bologna, I-40129, Italy\\
$^{2}$CSIRO--ATNF, P.O. Box 76, Epping, NSW 1710, Australia\\
$^{3}$INAF--IRA Bologna, Via Gobetti 101, Bologna, I-40129, Italy}
\begin{document}

\date{Accepted xx xx xx. Received yy yy yy; in original form zz zz zz}

\pagerange{\pageref{firstpage}--\pageref{lastpage}} \pubyear{2005}

\maketitle

\label{firstpage}

\begin{abstract}
Polarized diffuse emission observations at 2.3~GHz in a
high Galactic latitude area 
are presented. The $2\degr\times2\degr$ field, centred in 
($\alpha = 5^{\rm h}$,~$\delta = -49^{\circ}$), is located in the region
observed by the BOOMERanG experiment.
Our observations has been carried out with the Parkes
Radio telescope and represent
the highest frequency detection 
done to date in low emission areas.
Because of a weaker Faraday rotation action,
the high frequency allows an estimate of the Galactic synchrotron 
contamination of the Cosmic Microwave
Background Polarization (CMBP) that is more reliable than that 
done at 1.4~GHz.
We find that the angular power spectra of the $E$-- and $B$--modes
have slopes of $\beta_E = -1.46\pm0.14$ and $\beta_B = -1.87\pm0.22$,
indicating a flattening with respect to 1.4 GHz.
Extrapolated up to 32~GHz, the $E$--mode spectrum
is about 3 orders of magnitude lower than
that of the CMBP, allowing a clean detection even at this frequency.
The best improvement concerns the $B$--mode, for which our single-dish 
observations provide the first estimate of the contamination on 
angular scales close to the CMBP peak (about 2 degrees). 
We find that the CMBP $B$--mode should be stronger than synchrotron 
contamination at 90~GHz for models with $T/S > 0.01$. This low level
could move down to 60--70~GHz the optimal window for CMBP measures.
\end{abstract}


\section{Introduction}
The Cosmic Microwave Background Polarization (CMBP) is a powerful tool
for investigating the early Universe. 
For instance,
inflationary models lead to a well defined peak--pattern of the
$E$--mode power spectrum and its precise measurement
provides a check of the inflationary paradigm
itself \citep{ko99}.

The second and fainter CMBP component -- the $B$--mode -- 
is predicted to directly probe Inflation. 
In fact, the emission on a few degrees
scale is related to the amount of gravitational waves (GW) generated by the Inflation,
and allows measurement of its main parameters, such as the
energy density of the Universe when this event
occurred \citep{ka98}. 
On subdegree scales, instead, the $B$-mode is
contaminated by the $E$-mode via galaxy gravitational 
lensing, allowing though a way 
to determine the matter fluctuation power spectrum 
\citep{zaldarriaga98}.
The first steps into $E$-mode measurements has begun with 
the detections of the DASI, CAPMAP and CBI experiments 
\citep{leitch04,barkats04,readhead04}. However,
we are far from a complete characterization
and the $B$-mode is still elusive.

Several astrophysical sources of polarized emission in the foreground must
be considered as they could contaminate the tiny CMBP signal.
Among them, the diffuse synchrotron emission
of the Galaxy dominates at low frequencies and it is expected to be
the major contaminant up to 100~GHz. 
Recently, evidence for a significant anomalous dust emission,
competitive with the synchrotron up to $50$~GHz, has been found
(e.g. \citealt*{deoliveira04,fink04}).
Although potentially relevant  
in total intensity, the small polarization fraction 
predicted \citep{lazarian00} seems to leave synchrotron as the
leading polarized component.

The study of the synchrotron emission
is thus crucial for CMBP experiments. Besides the estimate of the
contamination level, it allows 
the tuning of foreground separation procedures to extract 
the cosmic signal 
(e.g. \citealt{tegmark00}) and 
the selection of optimal sky areas for observation.

Polarization measurements with large sky coverage has becoming 
available at frequencies up to 1.4~GHz 
(e.g. see \citealt{carretti05} and references therein 
for a review of the available data), 
although their sensitivities and angular resolutions do not fully cover 
the CMBP needs.
An ideal field for CMBP studies is the high Galactic latitude 
field ($b \sim -38^\circ$) in the region observed by the BOOMERanG 
experiment.  This field, positioned at 
$\alpha = 5^{\rm h}$, $\delta = -49^{\circ}$, 
has low levels of synchrotron emission and has been
selected as the target area for 
CMBP experiments (e.g. BaR-SPOrt, \citealt{cortiglioni03}, and
BOOMERanG-B2K, \citealt{masi02}).
The first measurement of the polarized synchrotron
emission in low-emission areas was reported by \citet{be03}
in this field at 1.4 GHz with the Australia Telescope
Compact Array (ATCA).  
The analysis of 
these observations has allowed the
first estimate of the synchrotron contamination on the CMBP spectra
at high latitude \citep{carretti05}.
However, these authors found that at 1.4 GHz the area 
is likely to be affected by Faraday Rotation (FR) 
effects, which could enhance the detected emission 
and modify the spectra by
steepening their slopes.  In particular, their analysis shows that, 
at 1.4~GHz this area is in an intermediate state between negligible 
and significant FR effects. 
Since FR has a square 
dependence on the wavelength ($\delta\varphi\propto\lambda^2$),
higher frequencies should 
allow observations of the region without 
significant effects. Finally, the 1.4~GHz data have an 
angular scale sensitivity range limited by the interferometric 
nature of the observations, and cover only the angular scales up to 
15~arcmin.

To overcome these constraints, we have conducted 
single dish observations of this area 
at 2.3~GHz with the Parkes Radio telescope.
These observations also allow us to extend 
the angular scale sensitivity toward degree scales, 
leading to a firmer assessment of the contamination
of the CMBP $B$--mode generated by GW, which peaks on 
$\sim 2^\circ$ scales.

The paper is organized as follows: in Section~\ref{obsSec} we present
details of the observations along with the discussion of the obtained
maps.  In Section~\ref{specSec} we show the analysis of the angular
power spectra, while in Section~\ref{discSec} we discuss the
implications for CMBP measurements.

\section{Observations}\label{obsSec}

\begin{figure*}
  \includegraphics[angle=0, width=0.4\hsize]{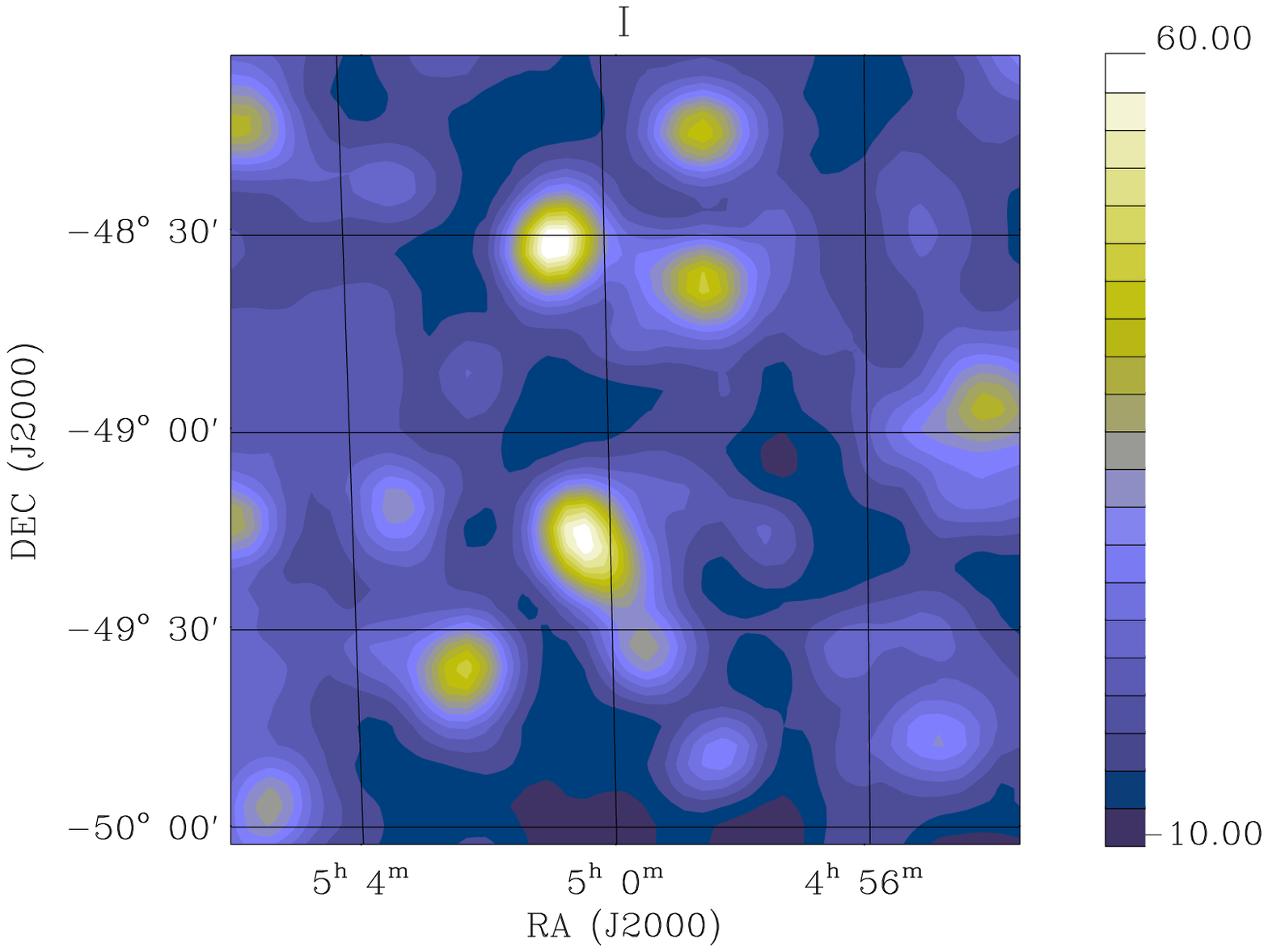}
  \includegraphics[angle=0, width=0.4\hsize]{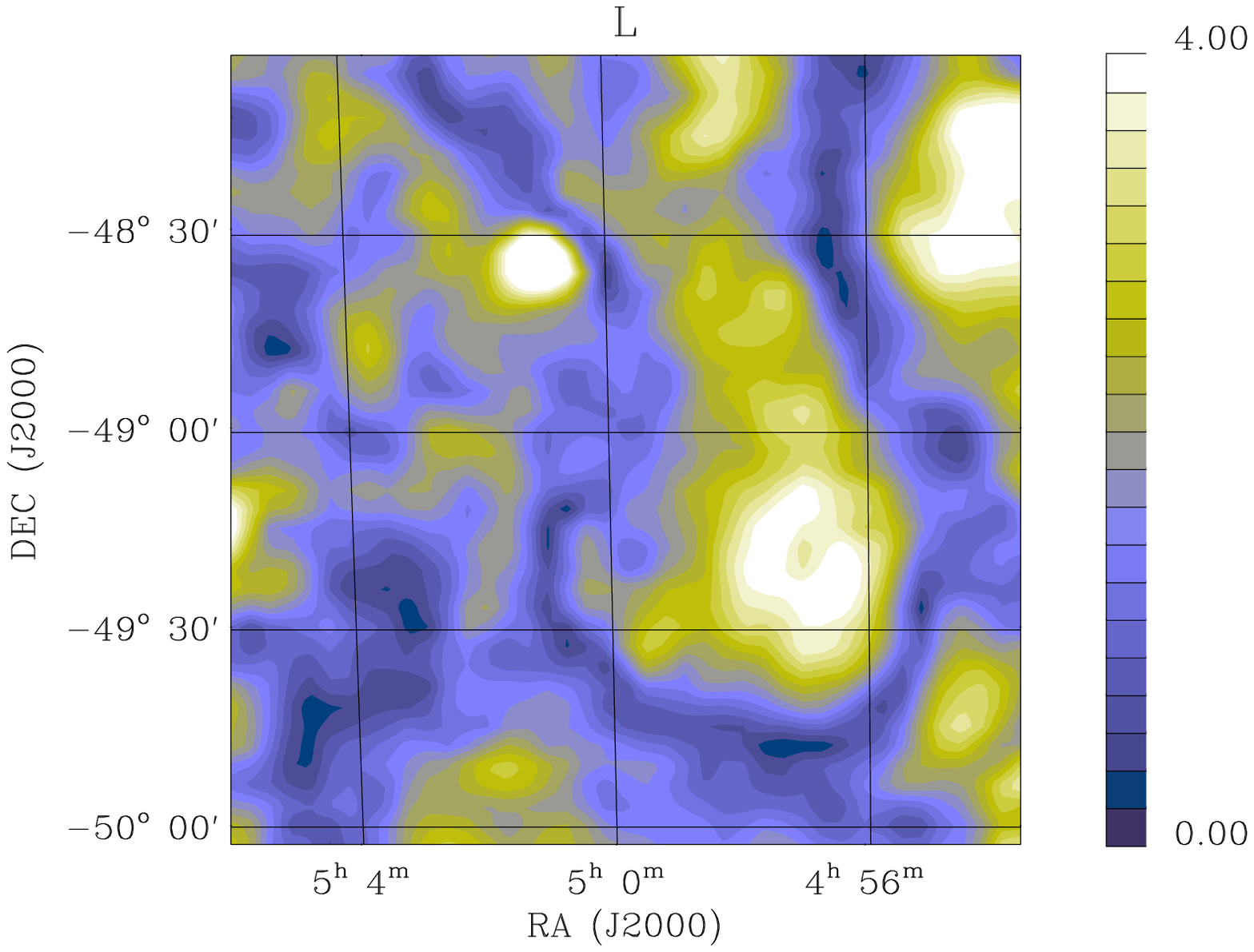}
  \includegraphics[angle=0, width=0.4\hsize]{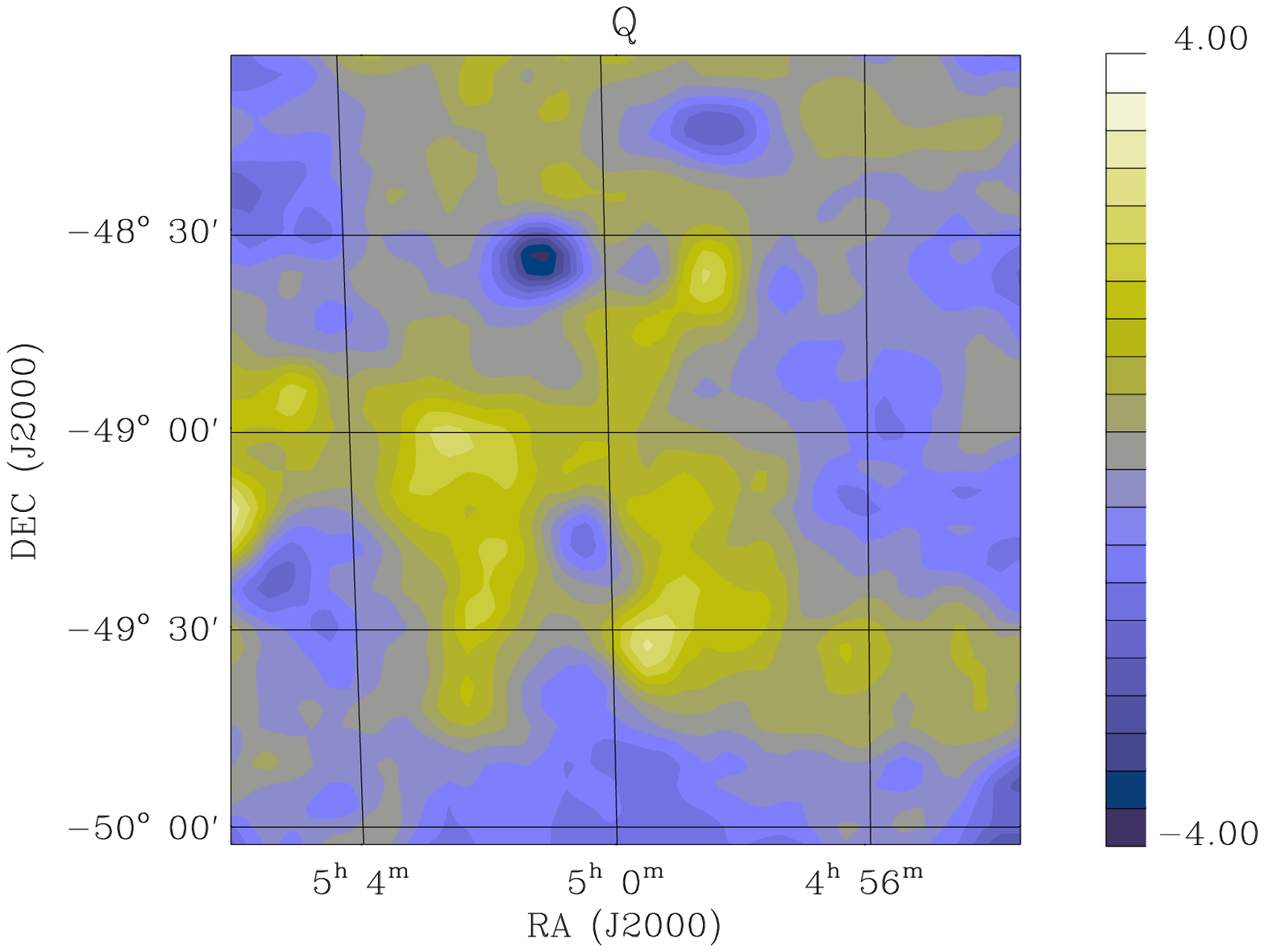}
  \includegraphics[angle=0, width=0.4\hsize]{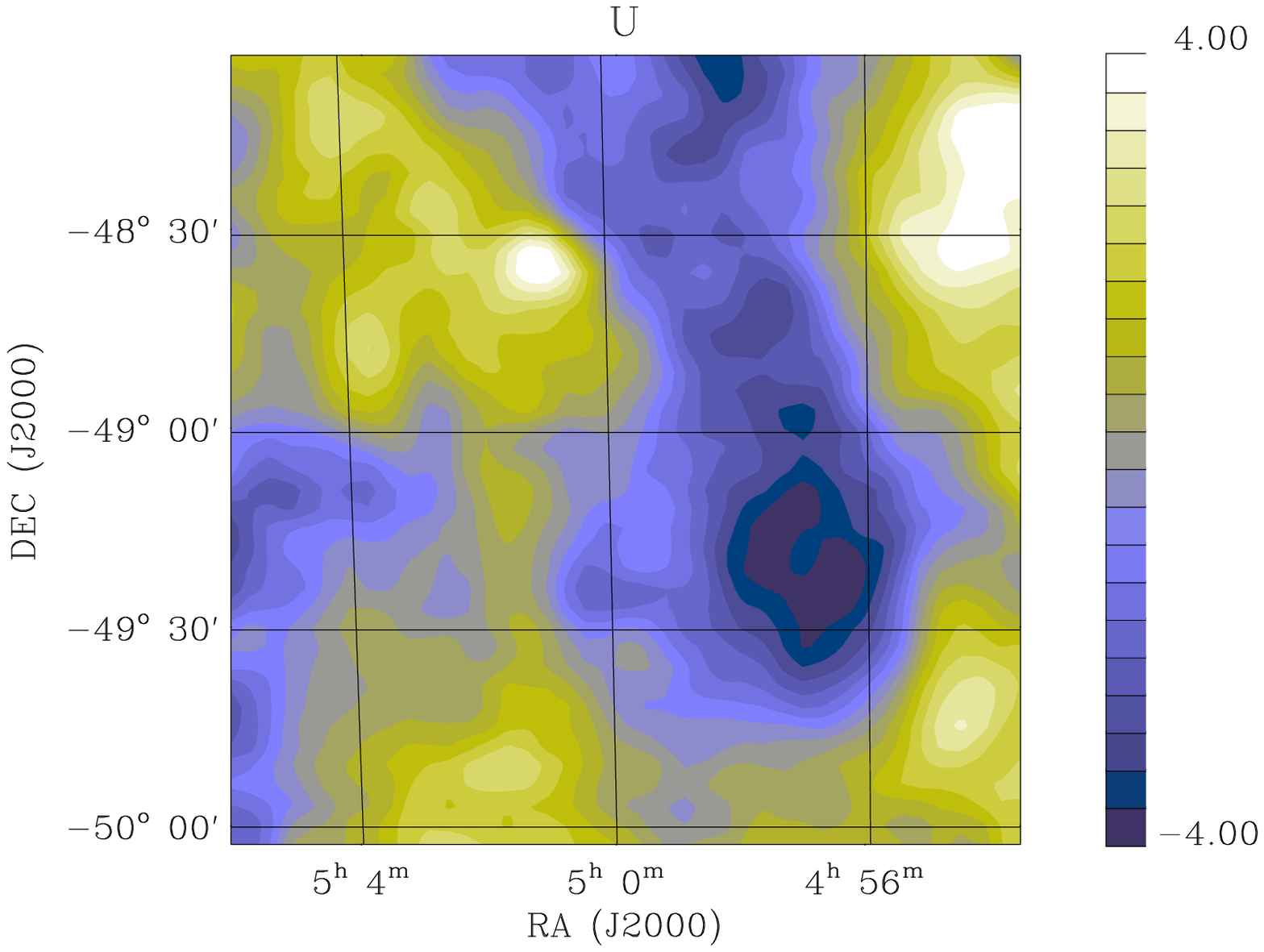}
\caption{Maps of the Stokes parameters $I$ (top-left), 
         $Q$ (bottom-left) and $U$ (bottom-right) of the observed area
         at 2.3~GHz. The polarized intensity $L=\sqrt{Q^2+U^2}$ is shown as well
         (top-right). Units are mK.\label{mapFig}
}
\end{figure*}

We have made single dish observations of the 
$2^\circ \times 2^\circ$ area centred in 
$\alpha = 5^{\rm h}00^{\rm m}00^{\rm s}$,~$\delta = -49^{\circ} 00^\prime 00^{\prime\prime}$ 
(J2000) ($l\sim255^\circ$,~$b\sim-38^\circ$) 
with the Parkes Radio telescope at 2.3~GHz on 17--23 August 2004.  
The field was surveyed with
$2\degr$ scans in both RA and Dec.  Scans were separated by 3~arcmin, 
providing an adequate sampling of the $8\farcm8$ telescope beam.  
We used the dual circular polarization Galileo receiver
in combination with a wide band 13~cm feed. 
The two total intensity channels and the two linear Stokes parameters,
$Q$ and $U$, were formed from the auto and cross correlations of the two
receiver outputs.
The digital correlator was configured to provide 
1024 250~kHz-channels, giving a total bandwidth of 256 MHz centered on
2300 MHz.
In the subsequent analysis 
the channels have been grouped in 8 $\times$ 32~MHz.
The system showed an optimal sensitivity in a 200~MHz wide band. After the
selection for RF interferences, a total 
effective bandwidth of $\Delta\nu = 128$~MHz has remained with a
central frequency of 2332~MHz.

The sources 1934-638 and 3C138 have been used for total intensity and
polarization calibration, respectively.  
The absolute polarization state of 3C138 was determined using
ATCA one week prior to the Parkes
observations.  Although 3C138 is variable, the timescale is such that
the one week lag between observations is insignificant \citep{padrielli87}.

Images of the Stokes parameters $I$, $Q$ and $U$ were constructed
through an iterative map--making procedure, 
based on the estimation and removal of a linear 
baseline from each scan. The
image produced at each iteration is used in the 
next to subtract the sky signal from the data and
improves the baseline evaluation (see \citealt{sbarra03} 
for the basic equations and Carretti \& Poppi, in preparation, 
for an implementation at the Medicina Radio telescope).
Differing from \citet{sbarra03}, a linear behaviour for the 
baseline is allowed. 
As usual, the removal of a linear baseline leads to the loss of the
absolute emission levels in the images.  

The Stokes $I$, $Q$, $U$ and linearly polarized intensity, 
$L=\sqrt{Q^2 + U^2}$, images
are shown in Figure~\ref{mapFig}.  The images cover an area of 
$2\degr \times 2\degr$, have a pixel size of 3~arcmin, 
and have an rms pixel sensitivity of 1.0~mJy~beam$^{-1}$ or 800~$\mu$K
in brightness temperature units (the gain is 0.77~Jy/K). 
Finally, they have been smoothed to a FWHM of $10\farcm 6$.

The total intensity emission is dominated by point sources, 
corresponding to the strongest sources in the 1.4~GHz image
\citep{carretti05}.
The polarization images are dominated by diffuse emission.
Structures are present on all angular scales up to the field size,
with the largest feature extending from north to south.
Excluding the one evident point source, the polarized intensity image
peaks at about 5~mK, while the peak-to-peak variation
of $U$ is of about 10~mK ($Q$ shows less power).

The small range of overlapping angular scales in the 1.4 and 2.3 GHz
observations makes
a direct comparison difficult.
However, it is worth noting the filamentary structure about 1~degree long
present at 1.4~GHz has no counterpart at 2.3~GHz 
but approximately
corresponds with the maximum gradient of the 2.3~GHz $U$ map.
It is possible that this structure is generated by a Faraday
screen at 1.4~GHz, as supposed also by \citet{be03}, but the structure
disappears at 2.3~GHz, where FR is weaker.

\begin{figure}
\centering
\includegraphics[angle=0, width=0.85\hsize]{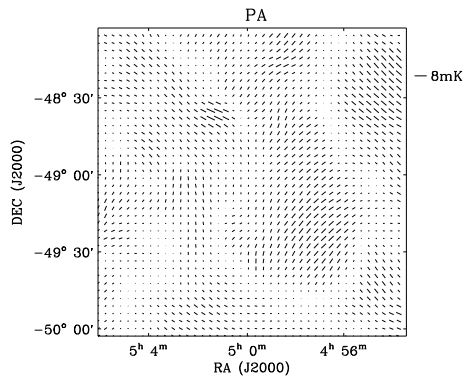}
\caption{Polarization angle map. The vector length is proportional to the 
         polarized intensity.\label{paFig}
}
\end{figure}

An image of the polarization angle is shown in Figure~\ref{paFig}.
The pattern is
uniform inside the large structures 
but sudden changes of about $90\degr$ are observed when crossing
a zero in polarized intensity.
These features can have physical origins, such as the effects of a
Faraday screen or a sudden change in the magnetic field direction.
However, here they correspond to a change of sign
of $U$ (the dominant component) 
and are characterized by a
$\sim 90^\circ$ change of polarization angle.  
This could be simply generated by a mean value removal. 
The addition of a constant value to $U$ would eliminate this
change of sign, keeping more uniform the polarization angle pattern.
Map-making procedures can generate such a situation, 
therefore, besides the presence of either a
Faraday screen or a rapid change of the magnetic field, the sudden
changes in polarization angle of our map can be related to this
non-physical cause.  
Only observations of a larger area can provide the missing mean 
emission and distinguish the two cases.
\begin{figure}
\centering
\includegraphics[angle=0, width=0.85\hsize]{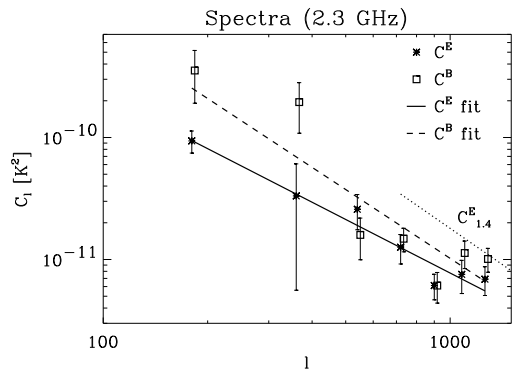}
\caption{Angular power spectra $C^E_\ell$
and $C^B_\ell$ in the observed area together with their power law fits.
$C^B_\ell$ is slightly shifted to the right to avoid confusion.
The fit to the 1.4~GHz $E$--mode power spectrum scaled up to 2.332~GHz 
($C_{1.4}^E$) is also displayed 
(a frequency spectral slope of $\alpha = -2.8$ is assumed).
\label{specFig}}
\end{figure}
\begin{figure}
\centering
\includegraphics[angle=0, width=0.85\hsize]{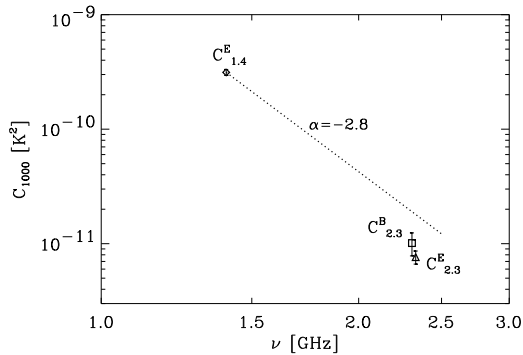}
\caption{Fits to power spectra $C^X_{\ell}$ computed at $\ell = 1000$
for both 1.4~GHz and 2.3~GHz. $E$-- and $B$--mode are almost overlapped at 1.4~GHz, 
so that the latter has been omitted to avoid confusion. 
Similarly, $C^B_{2.3}$ is slightly shifted left to avoid confusion.
The dotted line represents the 1.4~GHz value extrapolated 
using a brightness temperature slope of $\alpha = -2.8$.
\label{spec2Fig}}
\end{figure}

\section{Power Spectrum Analysis}
\label{specSec}
We compute the power spectrum of the $E$-- and $B$--modes of the
polarized component through the Fourier
technique of \citet{seljak97}.  The results are shown in
Figure~\ref{specFig} together with power law fits
to the equation:
\begin{equation}
    C_{\ell}^X = C_{500}^{X} \left({\ell \over 500}\right)^{\beta_X},
    {\rm with\;\;} X=E,\,B,
\end{equation}
where $\ell$ is the multipole, corresponding to
the angular scale $\theta \approx 180^\circ / \ell$.
The spectra of $E$-- and $B$--mode have similar power, except at
smaller $\ell$, where the poor statistics allows larger deviations
between the two components.
The results of the fits are given in Table~\ref{specTab}.
\begin{table}
 \centering
  \caption{Fit parameters for $E$ and $B$ spectra.\label{specTab}}
  \begin{tabular}{@{}lcc@{}}
  \hline
   Spectrum     &   $C_{500}^X$ [$10^{-12}$~K$^2$]     &   $\beta_X$        \\
  \hline
  $C^E_\ell$ &  $21 \pm 2 $  &  $-1.46 \pm 0.14$ \\
  $C^B_\ell$ &  $37 \pm 6 $  &  $-1.87 \pm 0.22$ \\
  \hline
  \end{tabular}
 \label{powFitTab}
\end{table}

The comparison with the spectra obtained in the same area at 1.4~GHz
\citep{carretti05}
allows interesting considerations.
Figure~\ref{specFig} plots the 1.4~GHz spectrum scaled to 
2.3~GHz with a brightness temperature frequency spectral
slope $\alpha = -2.8$, 
while Figure~\ref{spec2Fig} shows the $E$--mode 
powers for $\ell=1000$ at 1.4 and 2.3~GHz.
The $E$--mode emission level of our 2.3~GHz observations is 
significantly weaker than expected for a synchrotron spectrum
with slope of $\alpha = -2.8$.
In fact, a frequency spectral index of
$\tilde{\alpha}_E = -3.6 \pm 0.15$ is
required to fit the amplitudes. A similar view is given
by the $B$--mode, although the needed slope is flatter
($\tilde{\alpha}_B = -3.3 \pm 0.2$).
Such steep slopes are unlikely at these frequencies for synchrotron
emission \citep{platania98}.  We suggest that the amplitudes at 1.4~GHz are
affected by FR.
This suggestion is supported by \citet{carretti05}, who find that
randomization of polarization angles induced by FR
can transfer power from large to small angular scales, enhancing
the power on subdegree scales.  At 2.3~GHz, where Faraday effects are
less than at 1.4~GHz, the observed polarization is more closely related
to the intrinsic emission, as discussed in \citet{carretti05}.
Thus it seems clear that observations at higher frequencies
are more robust for CMB extrapolation.

The comparison between the slopes of the angular power spectra
provides a similar view. At 2.3~GHz $\beta_E$ is flatter than
at 1.4~GHz, where a value of $\beta_E^{1.4}\sim -1.97\pm 0.08$ has been
measured. The values measured at 2.3~GHz are closer to the mean value
$\beta_X^{GP} \sim -1.6$ ($X=E,B$) obtained on the Galactic plane
at the same frequency \citep{bruscoli02}.  
\citet{carretti05} discussed the
effects of the FR on the power spectrum slope,
finding that steepening can occur if FR action is
significant. In this context, the flatter slope
measured at 2.3~GHz can be interpreted as FR effects being
weak relative to those at 1.4~GHz.
We note that the $B$--mode cannot be fit well with a power law
(see Figure~\ref{specFig}), so that
the resulting larger error makes $\beta_B$ compatible with both
$\beta_B^{1.4} = -1.98 \pm 0.07$ and $\beta_X^{GP}$. Therefore, 
no firm statements can be made about the comparison of 
$\beta_B^{1.4}$ and $\beta_B$.

\section{Discussion}
\label{discSec}

The steep spectral index needed to match the 1.4~GHz emission level,
the regularity in the polarization angle pattern,  and the flatter
slope of the 2.3~GHz $E$--mode angular power spectrum, indicate that our
maps are weakly affected by FR effects. Consequently, 
they provide the most reliable estimates of the contamination of the CMBP
by synchrotron emission at high Galactic latitudes. Moreover, these data
explore larger angular scales than possible with the 1.4~GHz
observations (limited to scales smaller than 15$\arcmin$), allowing
estimates closer to the peak of the $B$--mode emission at about
$2^\circ$ ($\ell \sim 100$).
\begin{figure}
\centering
\includegraphics[angle=0, width=0.85\hsize]{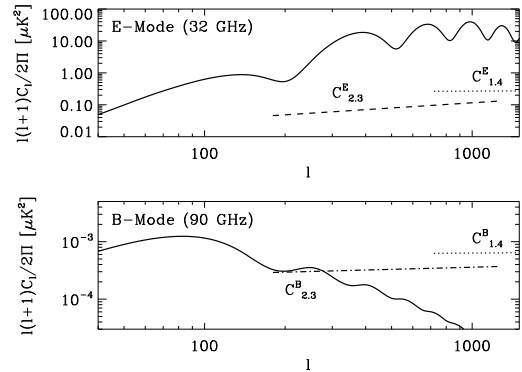}
\caption{Angular power spectra $C^E_\ell$ (top)
and $C^B_\ell$ (bottom)
extrapolated to 32~GHz and 90~GHz, respectively, 
of both the 2.3 and 1.4~GHz data.
A frequency spectral slope of $\alpha = -3.1$ is assumed.
Spectra expected for the CMBP are shown for comparison (solid) by using
cosmological parameters determined with the WMAP data
\citep{spergel03}. For the $B$--mode, a tensor-to-scalar power ratio
$T/S = 0.01$ is  assumed.\label{specCMBFig}}
\end{figure}

The extrapolations up to 32~GHz ($E$--mode) and 90~GHz ($B$--mode)  are
shown in Figure~\ref{specCMBFig}, where the typical spectral index
$\alpha = -3.1$ of the 1.4--23~GHz range has been assumed
\citep{bernardi04}. 
The contamination of the $E$--mode is significantly
lower than that estimated with the 1.4~GHz data: a factor two better is
seen at $\ell\sim 1000$, while the flatter slope of the 2.3~GHz angular
spectrum increases that factor at larger scales to about~4 at 
$\ell = 400$. 
Note that the  $\ell$-range now covered allows a direct estimate
on angular scales usually probed by 30~GHz experiments (e.g. the
32~GHz channel of BaR-SPOrt with a FWHM~=~$0.4^\circ$).
This improves the expected detectability of the CMBP $E$--mode signal,
by avoiding the uncertainties of an angular extrapolation.  
At 32~GHz, the first CMBP peak  ($\ell\sim 400$) is about 3 orders 
of magnitude above the
synchrotron emission (i.e. a factor 30 in signal). This makes us
confident that in this area the cosmic signal is detectable by
experiments in this band. 
Even considering an uncertainty 
of $\Delta\alpha=0.2$ in the
spectral index \citep{platania98,bennett03}, the extrapolation would
change by a factor $\sim 3$ in spectrum, not significantly affecting 
the previous conclusion. 
In addition, our estimate strengthens
the results obtained by the CMBP experiments, which
find indications that at 30 GHz the emission of this 
foreground is not dominant \citep{leitch04,readhead04}.
At 90 GHz the frame is even better: the
synchrotron $E$-mode is 5~orders of magnitude weaker than the CMB,
giving a very negligible contamination in the CMB frequency window.

Our measurements allow even larger improvements in the expected
detectability of the $B$--mode component.
The improved coverage of the
angular spectrum gives a more reliable estimate of the $B$--mode
contamination near the  $\ell\sim 100$ CMBP peak.  
The power of this peak is expected to vary with the 
tensor-to-scalar perturbation ratio
$T/S$ so measuring the level of the GW
background in the early Universe. 
Our data suggest that at 90~GHz the synchrotron contamination 
dominates the CMBP signal for models with  $T/S < 0.01$,
allowing the detection of this still unknown parameter well below
its present upper limit ($T/S < 0.90$ -- 95\% C.L.:
\citealt{spergel03}).

The resulting picture is encouraging:  
in this area of the sky the CMBP $E$--mode is expected a factor 30
larger than the synchrotron emission at $\sim 30$~GHz, 
so appearing free of this contaminat, 
even allowing for possible uncertainties in the frequency 
extrapolation.  
At 90~GHz the contamination on
the $E$--mode is negligible, while that on
the $B$--mode appears to dominate the CMB signal only for
models with $T/S < 0.01$.
This low level is beyond the capability of the the PLANCK mission 
($\Delta\,(T/S)\sim 0.06$ -- 95\%~C.L.: see \citealt{tegmark00}) 
and only future experiments devoted to the search of the $B$-mode will
be limited by the foreground emission in this sky area.
\begin{figure}
\centering
\includegraphics[angle=0, width=0.90\hsize]{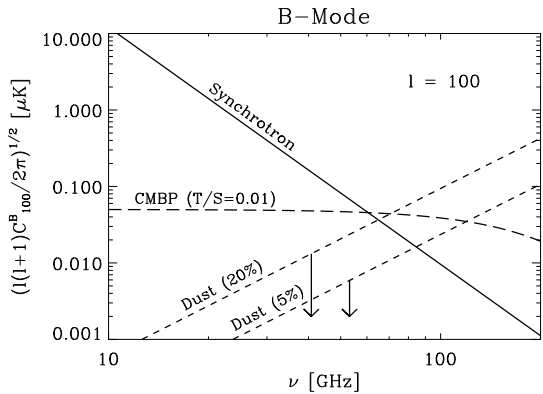}
\caption{Frequency behaviour of the synchrotron
and dust $B$--mode in our patch. The quantity
$\sqrt{\ell (\ell + 1) C^B_l / (2 \pi)}$ in $\ell = 100$ is shown
because a fair estimate of the signal on the scale where the CMB
$B$--mode peaks ($\theta \sim 2^\circ$).
The CMB level expected in the case of $T/S = 0.01$ is  plotted
for comparison.
Synchrotron emission is estimated by extrapolating to $\ell = 100$
the fit of Table~\ref{specTab}.
We assume a slope of $\alpha = -3.1$ for the frequency behaviour.
Dust emission is evaluated from the total intensity
upper limit measured with the BOOMERanG experiment at $b=-38^\circ$
\citep{masi01}. A slope of $\alpha = 2.2$
is assumed for the frequency extrapolation \citep{bennett03}.
Two values of polarization fraction (5\% and 20\%)
are used to bracket the 10\% deduced by \citet{benoit04}
for high Galactic latitudes.}
\label{foregFig}
\end{figure}

Although an accurate estimate of the dust
contribution in the area cannot be done due to the many uncertainties,
we indicate some plausible upper limits in Figure~\ref{foregFig}.
These, and our new lower estimate of the synchrotron component, 
suggest that the thermal dust could become the most important source of
foreground  noise at 90~GHz. This would move the optimal window for
$B$--mode investigations from 90--100~GHz toward 70~GHz,
as for the CMB anisotropy \citep{bennett03}.
Better estimates of the optimum $B$--mode observing frequency require a
direct measure of the polarized dust emission in this sky area, 
possibly at higher frequency where the dust
emission is stronger.

\section*{Acknowledgments}

This work has been carried out in the frame of the SPOrt experiment, 
a program funded by ASI (Italian Space Agency).
G.B. and S.P. acknowledge ASI grants.
We wish to thank Bob Sault for 
the ATCA observations of 3C138, 
Warwick Wilson for his support in the
Wide Band Correlator set-up, and
John Reynolds for his support in the receiver set-up.
Part of this work is based on observations taken with
the Parkes Radio telescope, which is part of the Australia Telescope,
funded by the Commonwealth of Australia for operation
as a National Facility managed by CSIRO.
We acknowledge the use of the CMBFAST package.

\bsp

\label{lastpage}


\begin{thebibliography}{99}
\bibitem[\protect\citeauthoryear{Barkats et al.}{2004}]{barkats04} Barkats D., et al.,
        2004, ApJ, 619, L127
\bibitem[\protect\citeauthoryear{Bennett et al.}{2003}]{bennett03} 
        Bennett C.L., et al., 2003, ApJS, 148, 97
\bibitem[\protect\citeauthoryear{Beno\^{\i}t et al.}{2004}]{benoit04} 
        Beno\^{\i}t A., et al., 2004, A\&A, 424, 571
\bibitem[\protect\citeauthoryear{Bernardi et al.}{2003}]{be03} Bernardi G.,
        Carretti E., Cortiglioni~S., Sault~R.J., Kesteven~M.J., Poppi~S., 
        2003, ApJ, 594, L5
\bibitem[\protect\citeauthoryear{Bernardi et al.}{2004}]{bernardi04} Bernardi G.,
        Carretti E., Fabbri~R., Sbarra~C., Cortiglioni~S., Poppi~S., Jonas~J.L.,
        2004, MNRAS, 351, 436
\bibitem[\protect\citeauthoryear{Bruscoli et al.}{2002}]{bruscoli02} Bruscoli M., 
        Tucci M., Natale V., Carretti~E., Fabbri~R., Sbarra~C., Cortiglioni~S.,
        2002, New Astron., 7, 171
\bibitem[\protect\citeauthoryear{Carretti et al.}{2005}]{carretti05} Carretti E., 
        Bernardi G., Sault~R.J., Cortiglioni~S., Poppi~S., 2005, MNRAS, in press,
        astro-ph/0412598 
\bibitem[\protect\citeauthoryear{Cortiglioni et al.}{2003}]{cortiglioni03} Cortiglioni S., 
        et al., 2003, 
        in Warmbein~B., ed., 16th ESA Symposium on European Rocket and Balloon Programmes 
        and Related Research, ESA Proc. SP-530, p. 271	
\bibitem[\protect\citeauthoryear{de Oliveira--Costa et al.}{2004}]{deoliveira04} 
         de Oliveira--Costa A., Tegmark M., Davies~R.D., Guti\'errez~C.M., Lasenby~A.N.,
         Rebolo R., Watson R.A., 2004, ApJ, 606, L89
\bibitem[\protect\citeauthoryear{Finkbeiner, Langston \& Minter}
        {Finkbeiner et al.}{2004}]{fink04} Finkbeiner~D.P., Langston~G.I., 
        Minter~A.H., 2004, ApJ, 617, 350
\bibitem[\protect\citeauthoryear{Kamionkowski \& Kosowsky}{1998}]{ka98} 
        Kamionkowski~M., Kosowsky~A., 1998, PRD, 57, 685
\bibitem[\protect\citeauthoryear{Kosowsky}{1999}]{ko99} 
        Kosowsky A., 1999, New Astron. Rev., 43, 157
\bibitem[\protect\citeauthoryear{Lazarian \& Draine}{2000}]{lazarian00} 
        Lazarian A., \& Draine B.T., 2000, ApJ, 536, L15
\bibitem[\protect\citeauthoryear{Leitch et al.}{2004}]{leitch04} Leitch~E.M., 
        Kovac~J.M., Halverson~N.W., Carlstrom~J.E., Pryke~C., Smith~M.W.E.,
        2004, astro-ph/0409357
\bibitem[\protect\citeauthoryear{Masi et al.}{2001}]{masi01} 
        Masi S., et al., 2001, ApJ, 553, L93
\bibitem[\protect\citeauthoryear{Masi et al.}{2002}]{masi02} 
        Masi S., et al., 2002,
        in Cecchini~S., Cortiglioni~S., Sault~R.J., Sbarra~C., eds, 
        Astrophysical Polarized Backgrounds, AIP Conf. Proc., 609, 122
\bibitem[\protect\citeauthoryear{Padrielli et al.}{1987}]{padrielli87} 
        Padrielli L., et al., 1987, A\&AS, 67, 63
\bibitem[\protect\citeauthoryear{Platania et al.}{1998}]{platania98}	
        Platania~P., Bensadoun~M., Bersanelli~M., de~Amici~G., Kogut~A., 
        Levin~S., Maino~D., Smoot~G.F., 1998, ApJ, 505, 473
\bibitem[\protect\citeauthoryear{Readhead et al.}{2004}]{readhead04}
        Readhead A.C.S., et al., 2004, Science, 306, 836
\bibitem[\protect\citeauthoryear{Sbarra et al.}{2003}]{sbarra03}
        Sbarra~C., Carretti~E., Cortiglioni~S., Zannoni~M., Fabbri~R., 
        Macculi~C., Tucci~M., 2003, A\&A, 401, 1215
\bibitem[\protect\citeauthoryear{Seljak}{1997}]{seljak97} Seljak~U., 
        1997, ApJ, 482, 6
\bibitem[\protect\citeauthoryear{Spergel et al.}{2003}]{spergel03} Spergel~D.N.,
        et al., 2003, ApJS, 148, 175
\bibitem[\protect\citeauthoryear{Tegmark et al.}{2000}]{tegmark00} Tegmark~M.,
        Eisenstein~D.J., Hu~W., de Oliveira--Costa~A., 2000, ApJ, 530, 133
\bibitem[\protect\citeauthoryear{Zaldarriaga \& Seljak}{1998}]{zaldarriaga98} 
        Zaldarriaga~M., Seljak~U., 1998, PRD, 58, 23003
\end{thebibliography}
\end{document}